# Accelerating Socio-Technological Evolution: from ephemeralization and stigmergy to the global brain

Francis Heylighen

http://pespmc1.vub.ac.be/HEYL.html
ECCO, Vrije Universiteit Brussel

**Abstract:** Evolution is presented as a trial-and-error process that produces a progressive accumulation of knowledge. At the level of technology, this leads to ephemeralization, i.e. ever increasing productivity, or decreasing of the friction that normally dissipates resources. As a result, flows of matter, energy and information circulate ever more easily across the planet. This global connectivity increases the interactions between agents, and thus the possibilities for conflict. However, evolutionary progress also reduces social friction, via the creation of institutions. The emergence of such "mediators" is facilitated by stigmergy: the unintended collaboration between agents resulting from their actions on a shared environment. The Internet is a near ideal medium for stigmergic interaction. Quantitative stigmergy allows the web to learn from the activities of its users, thus becoming ever better at helping them to answer their queries. Qualitative stigmergy stimulates agents to collectively develop novel knowledge. Both mechanisms have direct analogues in the functioning of the human brain. This leads us to envision the future, super-intelligent web as a "global brain" for humanity. The feedback between social and technological advances leads to an extreme acceleration of innovation. An extrapolation of the corresponding hyperbolic growth model would forecast a singularity around 2040. This can be interpreted as the evolutionary transition to the Global Brain regime.

## Evolutionary progress

The present paper wishes to directly address the issue of globalization as an evolutionary process. As observed by Modelski (2007) in his introductory paper to this volume, globalization can be characterized by two complementary processes, both taking place at the planetary scale: (1) growing connectivity between people and nations; (2) the emergence of global institutions. The first process is essentially economical and technological: the flows of matter, energy and information that circulate across the globe become ever larger, faster and broader in reach, thanks to increasingly powerful technologies for transport and communication, which open up ever larger markets and forums for the exchange of goods and services. The second process is fundamentally political and social: these increasingly powerful flows that cross the national borders and therefore the boundaries of most jurisdictions need to be efficiently regulated. This requires the development of a complex, global system of agreements between all the actors involved, specifying the rules to be followed and the mechanisms to enforce them.

The present paper wishes to explore the deeper evolutionary forces driving these two processes. These forces are so fundamental that we find them not only in the evolution of global society, but in evolutionary processes at the physical, chemical, biological or cognitive level (Heylighen 1999a; 2007a; Stewart 2001). Although I will mostly discuss issues of globalization to illustrate these effects, the conceptual framework is perfectly general, and equally applicable to other domains, such as animal behavior. In agreement with (Devezas & Modelski 2003; Modelski 2007), I see evolution basically as a *learning process*. This implies a specific evolutionary dynamics with a non-arbitrary directionality.

However, unlike Modelski, Devezas and most other authors in this collection, I do not see any clear cyclicity in this process. Instead, I will argue that the effects of evolution can best be represented as a monotonous (always increasing) function, although the speed of the corresponding advance is variable, with ups and downs that may give an impression of periodicity, but—most importantly—a strong tendency towards acceleration. This combination of monotonicity and acceleration allows us to make extrapolations for the relatively short-term future. However, the trends I will be speaking about are mostly qualitative, and can only be partially captured by numerical data. Therefore, my forecast of globalization is basically a complex "picture" of how a future global social system may look like, which I will designate by the metaphor of the *global brain*.

The theoretical background that supports this model may be called evolutionary-cybernetic, since it integrates the main insights from cybernetics and from evolutionary theory (Heylighen 2007c; Turchin 1977). It is closely related to the presently popular approach of complex adaptive systems (Holland 1996) with its basis in multi-agent computer simulations, although its historical roots are more ancient. Cybernetics (Ashby 1964; Heylighen & Joslyn 2001) is the science that studies how goal-directed systems can succeed in a complex, variable environment, by counteracting any perturbations that make them deviate from their preferred course. Adopting a more modern terminology, we will call such systems that try to reach their goal by acting upon their environment *agents*. Agents can be people, organizations, cells, robots, or any living organisms.

Evolutionary theory adds that the implicit goal of all systems created through natural selection (and thus of all non-artificial agents) is the maximization of *fitness*, i.e. survival, reproduction and development. Evolution can in general be seen as a learning process, during which the evolving system accumulates knowledge or information about how best to survive and thrive in its environment. This has been argued by evolutionary epistemology (Campbell 1974), which sees all forms of knowledge as a product of evolution, and—in its more radical version—all evolutionary adaptation as a form of knowledge. Its main idea is that evolution is a problem-solving process based on trial-and-error, where the successful trials are selectively retained or "memorized"—thus adding to the evolving system's store of knowledge—, while the errors are eliminated.

Cybernetics adds a somewhat more abstract perspective to the basic evolutionary mechanism of variation and selection (Ashby 1962). Information as originally defined by Shannon is a reduction of uncertainty. Selection means the elimination of a number of possible variants or options, and therefore a reduction in uncertainty. *Natural selection therefore by definition creates information*: the selected system now "knows" which of its variants are fit, and which unfit. The more variation and selection it undergoes, the more knowledge or information it accumulates, and the better it will be able to tackle the problems that the environment confronts it with. To have variation, it is sufficient that the system undergoes some form of either random or directed change—if only because of thermal fluctuations. As Ashby (1962) has shown, any system whose dynamics is not completely reversible will undergo selection, by leaving the unstable or unfit states and preferentially seeking out the fitter ones (Gershenson & Heylighen 2003).

Therefore, all systems tend to evolve towards better adaptation or fit, which implies greater information or knowledge about their environment. In that sense, there is an unambiguous advance, "progress" or "arrow" characterizing evolution, which tends to be accompanied by greater complexity, intelligence and integration (Heylighen 1999a; Stewart 2001; Wright 2000). Note that this philosophy does not preclude errors, setbacks, or "overshoots": variation is a process that by definition makes errors, sometimes resulting in an overall reduction of fitness; moreover, the environment changes, and what used to be fit may no longer be fit under changed circumstances. However, considered in the longest term and the largest scale, evolution is clearly progressive (Heylighen 1999a; Heylighen & Bernheim 2000b), in spite of the "postmodern" critiques of the idea of progress in biological evolution (Gould 1996) and in society (Marx & Mazlish 1996).

Having sketched the mechanism of progress in the most general or abstract sense, I will now apply it more concretely to understand the evolution, first, of technology and its impact on global connectivity, second, of social organization, and its implementation at the global level. I will then examine the variations in speed of evolutionary progress, arguing for acceleration, and against periodicity.

## Technological evolution

Ephemeralization

Let us start by perhaps the simplest and most obvious form of progress: *doing more with less*. All living organisms are subject to physical constraints: they require a finite amount of matter, energy and time in order to perform any action, including the most simple activity necessary for survival. These physical resources are subject to conservation laws: you cannot create matter or energy out of nothing. However, the same matter or energy can be used in many different ways, some highly efficient, others utterly wasteful. Given the general scarcity of resources for which a growing number of agents compete, this entails a strong selective pressure: agents that through trial-and-error have learned to use their resources more efficiently will have a strong advantage over the others, and thus will be picked out by natural selection. This applies in particular to society. If an agent can achieve the same or more output (products, services, information...) while requiring less input (effort, time, resources...) then that agent has increased its power to reach its goals—whatever these goals are. This increase in power applies under all circumstances, since a reduced need for effort or resources will make the agent less dependent on the particular conditions under which it tries to reach its goals.

This entails a strong *selective pressure* on all evolutionary systems in society: whenever there is a competition between individuals, groups, institutions, technologies or—most generally—systems of action, then *ceteris paribus* the more productive one will win. Indeed, whatever criterion directs the competition (producing cars, providing information, selling soft drinks, making religious converts...), the competitor who can achieve more for the same amount of investment will have a robust advantage over the others. This means that whenever a new variant appears that is somehow more productive than its competitors, it tends to become dominant, and the others will have to follow suit, or be eliminated. Thus, as long as there is *variation* (appearance of new variants) and *selection* (elimination of the less successful variants), evolution will produce an on-going increase in efficiency or productivity (Heylighen & Bernheim 2000b). Following Buckminster Fuller (1969), we may call this process of constantly achieving more with less *ephemeralization*.

Since the development of modern science in the 17th and 18th centuries, and its application to technology leading to the industrial revolution, this evolution has accelerated spectacularly. Rather than having to wait for a chance discovery, new techniques are now being developed in a *systematic* way, using the sophisticated methods for modelling and testing that characterize science. Ephemeralization moreover is self-reinforcing: the greater efficiency of institutions and technologies not only leads to greater output of goods and services, but also to a faster rate of further innovation, as new ideas are generated, developed, tested and communicated with less effort, while ever more time and energy becomes available to invest in research and development. The results are staggering: culture, society and even the natural world are changing in all aspects, and this at a breakneck speed (Toffler 1970).

Some well-known examples may illustrate this accelerating change. Because of better techniques, such as irrigation, crop improvement, fertilizers, pesticides and harvesting machines, agricultural productivity has increased spectacularly over the past two centuries: both the area of land and amount of human labor needed to produce a given amount of food has been reduced to a mere fraction of what it was. As a result, the price of food in real terms has declined with 75% over the last half century (Goklany 2000). Over the same period, the fuel consumption of cars has decreased just as spectacularly, while their speed, safety and comfort have increased. More generally, the average speed of transport has been increasing over the past few centuries, with the effect that people and goods need a much shorter time to reach any far-away destination. In the 16th century, Magellan's ships needed more than two years to sail around the globe. In the 19th century, Jules Verne gave a detailed account of how to travel around the world in 80 days. In 1945, a plane could do this trip in two weeks. Present-day supersonic planes need less than a day.

Without doubt, the most spectacular efficiency gains have been made in the transmission and processing of information. In pre-industrial times, people communicated over long distance by letters, carried by couriers on horseback. Assuming that an average letter contained 10,000 bytes, and that a journey took one month, we can estimate the average speed of information transmission as 0.03 bits per second. In the 19th century, with the invention of the telegraph, assuming that it takes a little over two seconds to punch in the Morse code for one character, we get a transmission rate of 3 bits per second. The first data connections between computers in the 1960's would run at speeds of 300 bit per second, another dramatic improvement. Presently, the most basic modems reach some 60,000 bits per second. However, the most powerful long distance connections, using fiber optic cables, already transmit billions of bits of per second. In a mere 200 years, the speed of information transmission has increased some 100 billion times!

We see a similar explosive development of power in information processing, which follows the well-known law of Moore, according to which the speed of microprocessors doubles every 18 month, while their price halves. As a result, a single chip used in a present-day electronic toy may contain more computing power than was available in the whole world in 1960. Again, this is a beautiful illustration of ephemeralization, as more (processing) is achieved with less (time, materials).

Reduction of friction

The net result of the drive towards increasing efficiency is that matter, energy and information are processed and transported ever more easily throughout the social system. This can be seen as a reduction of *friction*. Normally, objects are difficult to move because friction creates a force opposing the movement, which dissipates energy, and thereby slows down movement, until standstill. *Noise* plays a similar role in information transmission: over imperfect lines, parts of the signal get lost on the way, until the message becomes uninterpretable.

Physically, friction can be seen as the force responsible for the dissipation of energy and the concomitant increase of entropy (disorder), as implied by the second law of thermodynamics. Entropy increase entails the loss of information, structure, and "free" energy, that is, energy available for performing further work. This energy must be replenished from outside sources, and therefore a system performing work requires a constant input of energy carrying resources. However, the second law only specifies that entropy must increase (or remain constant), but not how much entropy is actually produced. Different processes or systems will produce entropy to varying degrees. Ephemeralization can be seen most abstractly as a *reduction of entropy production*, meaning that inputs are processed with less dissipation of resources. The result is that, for a given input, a system's output will contain more usable energy and information, and less noise or waste.

This has a fundamental consequence for cause-and-effect chains. Every process, object, or organization can be seen as an input-output system, which produces a certain output in reaction to a given input (Mesarovic & Takahara 1975). Inputs and outputs can be material, energetic and/or informational, but they are necessarily connected by a causal relation, which maps input (cause) onto output (effect) according to a particular set of rules or dynamics that characterizes the system. Given these rules, the state of the system, and the cause or input, you can predict the effect or output. What friction affects is the *strength* of this cause-effect relationship. A high friction or high entropy relation is one in which a strong, distinct cause will produce not more than a weak, difficult to discern, effect.

Imagine a billiard-ball (system) being hit by a billiard-cue (input or cause). The kinetic energy of the hit will be transferred practically completely to the ball, making it move with a speed proportional to the momentum imparted by the cue (output or effect). Imagine now hitting with that same cue a ball made of soft clay. The kinetic energy of the impact (input) will be almost completely absorbed or dissipated by the clay, resulting in a barely perceptible movement of the ball (output). The hard, smooth billiard-ball is a low friction system, with a strong cause-effect relation. The soft, irregular ball of clay, on the other hand, is a high friction system, with a weak cause-effect relation.

Now imagine coupling different causal processes or input-output systems in a chain. The output of the first system provides the input to the next one, and so on. If all systems in the sequence would be frictionless (an extreme, unrealistic case), any input given to the first system would be transmitted without any loss of strength to all subsequent systems. If the systems have friction, though, each next output will be weaker than the previous one, until it has become so weak that it no longer has any discernible effect

Let us discuss a few examples of such causal chains. Imagine a long, straight row of billiard-balls, each ball a short distance from the next one. If you hit the first ball with your cue (cause), it will hit the second ball (effect), which will itself hit the third ball (second effect), and so on. Because of friction, energy is lost, and each next ball will move more slowly than the previous one, until the point where the ball stops before it has reached the next one in line: the causal chain has broken. If the balls, and the surface on which they move, are hard and smooth, friction will be small, and a good hit may bring a dozen balls in motion. If balls and surface are soft or irregular, on the other hand, the chain is likely to break after a single step.

For an example more relevant to society, consider food production. The initial inputs of the chain are water, nutrients and sunlight, the resources necessary to grow crops. The final output is the food consumed by people. In between there are several processing and transport stages, each accompanied by a loss of resources. For example, most of the water used for irrigation will be lost by evaporation and diffusion in the soil before it even reaches the plants. From all the plant tissue produced, a large part will be lost because it is eaten by pests, succumbs to diseases or drought, rots away during humid episodes, etc. More will be lost because of damage during harvesting and transport. Further losses occur during storage because of decay, rodents, etc. Processing the fruits or leaves to make them more tasty or edible, such as grinding, cooking, or mixing with other ingredients, will only lead to further loss. What the consumer finally eats constitutes only a tiny fraction of the resources that went into the process.

As we noted above, ephemeralization has led to a spectacular reduction in these losses. In primitive agricultural systems, such as are still being used in many African countries, the output per unit of area or of water is minimal, and in bad years, hardly any produce will reach the population, leading to wide-spread famines. Modern techniques are much more efficient. For example, advanced irrigation systems bring the water via tubes directly to the root of the plant, minimizing evaporation and dissipation, and use sophisticated sensors in the leaves to monitor how much water the plant needs at any moment, so that they can supply just the amount for optimal growth. The gain compared to traditional irrigation systems, where water runs in ditches between the fields, can be a hundredfold. Similar gains are achieved during all stages of the production and distribution process, virtually eliminating losses because of pests, decay, oxidation, etc., with the help of refrigeration, pasteurization, airtight enclosures, various conserving agents, etc.

A last example of the role of friction in causal chains will focus on information transmission. Imagine giving your neighbor a detailed account of something that happened in your neighborhood, such as an accident or a police arrest. Your neighbor tells the story to his aunt, who passes it on to her friend, who tells it to her hairdresser, and so on. It is clear that after a few of such oral, person-to-person transmissions, very few details of the original account will have been conserved, because of forgetting, omissions, simplifications, etc. Moreover, the story is likely to have accumulated a number of errors, because of misunderstandings, embellishments, exaggerations, mixing up with other stories, etc. In the end, the story is likely to be forgotten and to stop spreading, or, in the rare case that some elements have caught the public's imagination, continue to spread, but in a form that is barely recognizable compared to the original. In either case, hardly anything will remain of the initial message. A simple way to reduce such "friction" or "noise" in this chain of "Chinese whispers" is to write down the account and send it to your neighbor by electronic mail. The neighbor can then simply forward the original message to his aunt, who forwards it to her friend, and so on. Unless someone actively manipulates the text, no information will be lost, and the causal chain will extend for as long as people are willing to forward the message.

Vanishing physical constraints

A general effect of ephemeralization is that things that used to be scarce or difficult to obtain have become abundant. For example, in the developed countries, the problem with food is no longer scarcity but overabundance, as people need to limit their consumption of calories in order to avoid overweight. Even in the poorest countries, the percentage of people that are undernourished is constantly decreasing (Goklany 2000; Simon 1995). More generally, the trend is clearly visible in the spectacular growth in wealth, usually measured as GDP per capita, since the beginning of the 19th century (Goklany 2000). The ever-increasing productivity not only results in people earning more, but in them needing to work fewer hours to achieve this wealth. Moreover, this economic development is typically accompanied by a general increase in the factors that underlie overall quality of life: health, safety, education, democracy and freedom (Heylighen & Bernheim 2000a; Goklany 2000; Simon 1995).

This is of course not to say that we live in the best of possible worlds. Many things are still much less abundant than we would like them to be, and although increasing productivity leads to an ever more efficient use of natural resources, ecologists have rightly pointed out that our present usage of many resources is unsustainable. The focus of this paper, though, is not on the remaining scarcities and wastages, which ephemeralization hopefully will sooner or later eradicate, but on the problems of social coordination created by such "hyperefficient" processes. To get there, we first need to understand more fundamentally how ephemeralization affects the dynamics of society.

In practice, most of the physical constraints that used to govern space, time, matter, energy and information have vanished. In the developed world most people can basically get as many material goods and as much information as they need, and this for a negligible investment in time and energy. (Of course, you can always *desire* more than you may need or be able to get). Moreover, distance as a factor has become largely irrelevant, as it costs hardly more effort to get goods, services or information from thousands of miles away than from a neighboring quarter. This is the real force behind globalization: the observation that social, economical and cultural processes no longer are impeded by geographical borders or distances, but cover the world as a whole. This is most clear on the Internet, where you can exchange information virtually instantaneously, without being aware whether your correspondent is situated around the corner, or on the other side of the planet. This practical disappearance of distance constraints has been referred to as *the death of distance* (Cairncross 2001), or *the end of geography* (O'Brien 1992).

Similarly, most of the constraints of duration have disappeared: apart from large-scale developments (such as building a house), most of the things an individual might need can be gotten in an interval of seconds (information, communication) to hours (most consumer goods and services). (In the Middle Ages, on the other hand, most of these commodities might have demanded months to acquire them, if available at all). Just imagine that you sit at your desk and suddenly start feeling hungry: a single phone call or web visit is sufficient to order a pizza, which will be delivered at your door 15 minutes later. The result may be called the *real-time society*: soon, all your wishes will be fulfilled virtually instantaneously, with a negligible waiting time.

Energy too is no longer a real constraint on the individual level: practically any system that we might need to produce some work or heat can just be plugged into the ubiquitous electricity network, to get all the energy it needs, for a price that is a mere fraction of our income. Finally, matter too becomes less and less of a constraint in any practical problem-solving. The raw material out of which a good is made (e.g. steel, plastic, aluminum) contributes ever less to the value of that good. In spite of dire warnings about the exhaustion of limited reserves, the real price of physical resources (e.g. copper, tin, coal, ...) has been constantly decreasing over the past century (Simon 1995), and has become insignificant as a fraction of the income we spend on consumption. This has led to a post-industrial economy that is mostly based on the exchange of immaterial resources, such as knowledge, organization and human attention. It is on this level, as we will see, that new, "cybernetic" issues emerge.

**Social evolution**

We have reviewed how the dynamics of evolution pushes agents to adopt ever more efficient methods and technologies, resulting in a minimization of physical and informational friction, and the virtual disappearance of the constraints of space, time and matter. But a similar dynamics affects social interactions between agents. Initially, agents are selected to be "selfish", i.e. to care only for their own benefit or "fitness", with a disregard for others except immediate kin (Dawkins 1989; Heylighen 2007a). But as the causal effects of their actions extend further over time and space, agents inevitably come to interact with an increasing number of other agents.

Social friction and the evolution of cooperation

Initially, interactions tend to be primordially competitive, in that a resource consumed by one agent is no longer available for another one. In that respect, interactions are characterized by *social friction* (Gershenson 2007), since the actions of one agent towards its goals tend to hinder other agents in reaching their goals, thus reducing the productivity of all agents' actions. Note that the two common meanings of the word "friction"—(physical) resistance, and (social) conflict—describe the same process of unintended obstruction of one process or system by another, resulting in the waste of resources. Even the actual mechanisms are similar, as illustrated by Helbing's (1992) mathematical model of the flow of pedestrians going into different directions, and thus unintentionally hindering each other's movements, in the same way that molecules in a fluid collide with other molecules, thus producing physical friction.

Like physical friction, social friction creates a selective pressure for reducing it, by shifting the agents' rules of action towards interactions that minimally obstruct other agents. Interactions, however, do not only produce friction, resulting in a loss of resources, they can also produce *synergy*, resulting in a gain of resources. Actions are defined to be synergetic if they produce more benefit when performed together than when performed separately. For example, a pack of wolves can kill larger prey when acting as a group than when acting on their own. These are the well-known advantages of cooperation (Dawkins 1989; Heylighen & Campbell 1995; Heylighen 2007a).

The evolution of cooperation is a complex and extensively researched subject (e.g. Axelrod 1984). The problem of overcoming the conflicts intrinsic to competitive relations is exemplified by the Prisoners' Dilemma and the Tragedy of the Commons (Heylighen & Campbell 1995). In different situations, different solutions have typically evolved. However, these solutions are all related, in the sense that they can be viewed as *institutions* in the broadest sense of the word, i.e. as socially agreed-upon systems of rules and control mechanisms for enforcing them, that regulate and coordinate interactions between agents so as to minimize friction and maximize synergy (Martens 2004; Wright 2000; Stewart 2000). Traffic rules provide a concrete example to illustrate the power of even the simplest institutions. Vehicles on the road compete for space. If two cars coming from different directions try to pass the same narrow crossing, they may obstruct each other to the point that neither of them can reach its destination. However, simple priority rules—if necessary supported by stop signs or traffic lights (Gershenson 2007)—can virtually eliminate this form of friction, letting everybody pass with a minimum of delay.

Recently, in collaboration with Carlos Gershenson and other PhD students of mine, I have started conceptualizing the spontaneous evolution or self-organization of such institutions as the emergence of a *mediator* (Gershenson 2007; Heylighen 2007a). Actions by definition affect the agent's environment. Insofar that agents share the same environment, the action of the one will have an effect on the situation of the other. This effect may be positive (synergy), negative (friction), or neutral (indifference). Therefore, the part of the environment that is shared (meaning that it is experienced by both agents) functions as a *medium* that carries their interactions. This medium affects, and is affected by, the agents. Agent and medium intimately interact and, therefore, co-evolve. (Although the medium is initially a purely passive, physical system, it too undergoes evolution, i.e. it experiences variations, some of which are selectively retained, some of which are eliminated).

The introduction of the variable "medium" in the equation allows us to avoid the classic Prisoners' Dilemma type of problems. Indeed, agent and medium are in general not competing, since they are wholly different types of entities requiring different types of resources (Heylighen 2007a). Therefore, it is easy for them to evolve a synergetic relation, i.e. such that the effect of both agent on medium and medium on agent are beneficial to the recipient. An example is an ecosystem in which a variety of species (agents) support their shared environment (medium) while being supported by it. Unsustainable interactions between agent and environment (medium) are just that: they cannot be maintained, and will eventually be eliminated by changes forced upon agent, medium or both. Thus, as always, natural selection in the long term produces increasingly stable or fit configurations.

## Stigmergy

Let us zoom in on the interaction between agents via the medium. An action by one agent that hinders another agent will tend to be resisted or counteracted by the second agent, with the result that the first agent fails to fully reach its goals. For example, a rabbit that tries to dig its hole near an ant nest will soon find its work undone by the activities of the ants. This creates a selective pressure for the agent to find a more effective action strategy, i.e. one that is unlikely to be obstructed. E.g., the rabbit will eventually give up, and choose another location for its burrow.

If the change in the medium brought about by this new action moreover happens to benefit another agent, that agent will tend to reinforce or support the change, thus in turn benefiting the first agent. Thus, there is a selective pressure on agents and their actions not only to reduce inter-agent friction, but also to promote synergy. For example, an animal that creates a passage across a field by flattening or breaking tall grasses will thus facilitate the movement of other animals, who will tend to follow in its footsteps, thus further flattening the trail, until a clear path is formed that is beneficial to all.

Note that this form of mutually beneficial interaction does not require any intention to cooperate, or even awareness of the other agent's existence. Such implicit collaboration, which was originally observed in social insects such as termites (Theraulaz & Bonabeau 1999), is called *stigmergy* (Susi & Ziemke 2001; Dorigo et al. 2000): the environmental change brought about by one agent's action incites another agent to act in turn, thus unconsciously contributing to their common benefit. This mechanism is general enough to explain the evolution of cooperation even in the absence of any form of rationality or ability to foresee the consequences of one's actions. This already allows us to side-step the Prisoners' Dilemma and other game-theoretical conundrums, which assume some form of rational decision-making from the agents. It moreover allows us to apply the procedure under the conditions of extreme unpredictability and complexity that characterize socio-technological evolution in this age of globalization.

The mechanism of stigmergy—i.e. indirect, environment-mediated cooperation—brings additional advantages. Because this form of action is directed at the shared environment, it will gradually reshape this medium into a structure that supports increasingly efficient and synergetic interactions. For example, the erosion of grasses, bushes and other obstacles along well-traveled routes will create a network of smooth paths connecting the most important destinations (such as feeding grounds and watering holes) for a group of animals. In places where several animals tend to pass at once, the path will widen so as to allow everyone free passage. This simultaneously reduces physical, informational and social friction: animals will be able to travel with less physical effort, less need to orient themselves, and less danger of obstruction from other animals.

After a while, the network of trails will have stabilized so that individual animals only need to contribute a minimal effort to its maintenance. Thus, the influence of individual agents on the medium tends to decrease. On the other hand, as the network becomes more reliable and extensive, the influence of the medium on the agents' activities increases. Eventually, the asymmetry is inverted: where initially we would see the agents as manipulating the medium, now it becomes more parsimonious to see the medium as directing the agents. The medium has turned into a *mediator*: it coordinates the individual activities so as to minimize friction or conflict, and to maximize synergy. The classic example of such an "active" coordination medium can be found in the pheromone trails that ants create while searching for food (Bonabeau et al. 1999; Dorigo et al. 2000). The trail network functions like an external memory or "collective mental map" for the ant colony, directing the individual ants to the different food sources and the nest via the most efficient routes (Heylighen 1999b). Similarly, in human society hiking paths and dirt roads eventually evolved into a dense network of streets and highways, complemented by road markings, separations between lanes, and traffic signs, that efficiently direct traffic so as to keep obstruction minimal.

Towards a global mediator

This leads us, after a perhaps long seeming digression, back to the evolution of our globalizing society. Human action has shaped a variety of media, i.e. shared environments supporting interaction. Initially, these were mostly concerned with the exchange of material goods and services. Examples are the different transport, industrial production and economic infrastructures. As discussed earlier, ephemeralization has made these increasingly efficient and global in reach. But there are limits to the reduction of physical friction: it becomes increasingly difficult to reduce the consumption of matter and energy simply because there is a minimal amount of matter/energy necessary for processes like feeding, movement, and construction. On the other hand, there is no clear limit to the reduction of informational or social friction, in that the losses of frictional interactions (negative sum) can be turned into the gains of synergetic interactions (positive sum) (Wright 2000; Heylighen & Campbell 1995).

One way to understand this unlimited capacity for growth is by noting that information, unlike matter and energy, is not a conserved quantity: it can in principle be replicated without limit. The Internet, which—because of its digital character, instantaneous communication, and negligible use of energy—can be viewed as a virtually frictionless medium, makes this unlimited replication possible in practice (Heylighen 2007b). As noted in the introduction, any evolutionary advance can be conceived as a gain in information. Communication between agents across one or more media makes it possible to spread that gain at an exponential rate. With a highly efficient, world-spanning medium like the Internet, a discovery made by one individual (say, a new way to avoid the flu) can in principle within days improve the life of people globally.

However, the remarkable efficiency of the Internet is at present still mostly physical or informational, not social. The Internet has grown so quickly that it has not had the time to evolve efficient institutions, i.e. collective systems of rules that coordinate individual actions. The result is a messy, confusing and constantly changing information landscape, that in principle offers immense benefits, but in practice only works reliably for a limited number of applications, while producing confusion, information overload, and various forms of "cybercrime". As a result, an individual discovery published on the net may indeed change the world's outlook within days, but the more likely outcome is that it will get buried within masses of other, mostly much less relevant information, and not receive the attention it deserves.

In conclusion, the evolutionary dynamics underlying globalization has already led to a relatively efficient physical distribution of matter, energy and information across the globe, but still needs to produce the social institutions that go with it. This is not a very original observation: critics of the globalization of markets have pointed out that the extension of the free trade in goods and services needs to be counterbalanced by the further development of transnational institutions, such as UN, UNESCO, WHO, etc., to protect the rights of children, workers, consumers, cultural groups or the environment (cf. Modelski 2007). The "stigmergic" theory proposed here, however, suggests a number of complementary mechanisms through which new types of institutions are likely to evolve.

The main idea is that the external interaction medium, a role that is increasingly dominated by the Internet, will evolve into a mediator. This mediator will not only facilitate, but direct, and eventually control, interactions so as to maximize their synergy. To achieve that, the medium needs to develop a form of intelligent management of the communication processes it supports, leading to what may be called *collective intelligence* (Lévy 1997; Heylighen 1999b) or *distributed cognition* (Susi & Ziemke 2001; Heylighen, Heath & Van Overwalle 2004). When this distributed intelligence spans the world, the resulting system may be called the *Global Brain* (Mayer-Kress & Barczys 1995; Goertzel 2001; Heylighen 2004, 2007c).

**The Emerging Global Brain**

As I have described both social and technological aspects and implications of the Global Brain concept in detail elsewhere (e.g. Heylighen 1999b 2004 2007c; Heylighen & Bollen 1996), I will here only present a short review, albeit from the new, stigmergic perspective. Two types of stigmergy can be distinguished (Théraulaz & Bonabeau 1999): quantitative and qualitative. Quantitative stigmergy *ranks* or *prioritizes* existing possibilities for action, thus helping agents to choose the action that is most likely to be beneficial. Qualitative stigmergy *creates* potential for action by changing the medium in such a way that novel possibilities arise.

### The web as neural network

The use of pheromones to mark foraging trails by ants is a paradigmatic example of quantitative stigmergy: the more often ants successfully travel a trail to find food, the more pheromones they leave behind, and therefore the more the trail becomes attractive to other ants searching for food. The strength of a pheromone trail is a quantitative measure of its probability to lead to a positive outcome. The basic mechanism whereby useful paths are gradually reinforced, and less useful ones weakened, provides a very general heuristic to tackle a variety of problems. It can be seen as a quantitative, stigmergic instantiation of the mechanism of evolution itself: maintain and grow the fit (useful); reduce, and eventually eliminate, the unfit. Under the label of "ant algorithms", it has become popular in computer science as a method to solve otherwise nearly intractable problems (Bonabeau et al. 1999; Dorigo et al. 2000).

Furthermore, the same mechanism seems to underlie learning in the brain: neuronal connections that are successfully used become stronger; the others become weaker. It is this analogy that initially inspired me to conceive of the World-Wide Web as a potential Global Brain (Heylighen & Bollen 1996). The web is a distributed network of documents connected by hyperlinks along which people travel ("surf") from page to page. The idea I developed together with my PhD student Johan Bollen was to strengthen and if necessary shortcut paths that are traveled frequently, while weakening the others, by applying a set of simple rules.

While the algorithm has not as yet been implemented on the scale of the web as a whole, a similar phenomenon already occurs implicitly: when people surfing the web end up in a particularly interesting page, they are likely to create one or more new links from their own pages pointing directly to it, thus shortcutting the long sequence they followed before finding it. This increases the number of links to the page, and the probability that other people would encounter it. The overall effect is captured by Google's PageRank algorithm, which provides a measure of the importance of a website as determined by the links directly or indirectly pointing to it. This is another example of quantitative stigmergy: the actions of many independent agents (people inserting links) on a shared medium (the web) produce a collective ranking (PageRank) that helps other agents find the options (documents) most likely to be useful to them.

As I have described elsewhere (Heylighen & Bollen 2002; Heylighen 1999b), several refinements can be conceived to make this mechanism much more efficient, thus enabling the web to rapidly learn from the way it is used and become ever better at anticipating and delivering what its users individually and collectively desire. This can potentially eliminate the friction caused by the chaotic organization of the present Internet, and the concomitant "data smog". Similar methods could be used to optimize not just data networks, but social networks. E.g. various services already exist on the web that introduce people to potentially useful contacts or partners (Coenen 2006).

Moreover, the optimized networks of priority-ranked trails that are created in this way can be traveled not only by humans, but by software agents. A "swarm" of such agents (Rodriguez 2007) is able to explore many paths in parallel according to the method of *spreading activation* (Heylighen & Bollen 2002)—which is again inspired by the functioning of the brain. It allows the intelligent network not only to explore a vastly larger array of potentially interesting information sources, but also to take into account the ever-changing context and often subconscious preferences of its users while selecting the sources most likely to be useful. The result is that users do not even need to enter keywords or explicitly formulate their queries, as their software agents implicitly learn their interests, while immediately taking into account changes in focus of attention.

Finally, agent swarms can perform not only the equivalent of intuitive, subconscious processes of activation spreading through the brain, but of systematic, logical search and deduction. To achieve that, the knowledge in the web needs to be organized according to a consensual ontology, i.e. a formal system of categories and relationships. Developing such ontologies is the goal of the Semantic Web project (Berners-Lee 1999). Given such a semantic network, a software agent could be programmed with a "grammar" of rules that tell it to only explore or return certain categories of nodes and links (Rodriguez 2007). A swarm of such agents should for example be able to find all birds that do not fly, or the most representative researchers (as measured e.g. by PageRank or citation impact) who have written about globalization and evolution, cite publications of Modelski, work in one of the NATO countries and have a PhD, so that you can invite them to your NATO-sponsored workshop.

In conclusion, quantitative stigmergy is able to turn the web from a passive medium for communication and storage of information into an intelligent mediator that uses learning and inference mechanisms similar to those of the human brain to recommend to its users the actions, information sources, or people most likely to be helpful for their aims. To achieve this, the intelligent web draws on the experience and knowledge of its users collectively, as externalized in the "trace" of preferences they leave on the paths they have traveled.

Collective production of new knowledge

Quantitative stigmergy as we have defined it only recommends the use of existing resources, it does not create anything new. To do that, we need to actively shape the medium into something that did not exist before. An example of such qualitative stigmergy in the world of social insects is nest-building by wasps, where individual wasps differentially add cells to the emerging structure of the nest (Théraulaz & Bonabeau 1999).

A more practical example is Wikipedia, the global electronic encyclopedia that is being written collaboratively by millions of people (Lih 2004). Any user of the web can add to or edit the text of any Wikipedia article—or create a new one, if its subject is not covered yet. All previous versions of an article are automatically stored, so that if an important section would be deleted by accident (or by intention), it can always be restored by a subsequent user. In that way, the information in Wikipedia can only grow, as the people who consult it add their own knowledge so as to improve the coverage of the subjects they are interested in. The activity is clearly synergetic since no single individual would be able to provide such an extensive coverage of all of humanity's knowledge. And since the different contributions are integrated into a well-organized and extensively cross-linked web of articles, the whole is clearly more than the sum of its parts.

Yet, the collaboration between Wikipedia contributors is essentially indirect. Over its history of a few years a typical article has been edited by a few dozen different people from different parts of the globe. In general, these people have never met or even communicated from person to person. Their only interaction is indirect, through the changes that the one makes to the text written by the other. When they disagree about how to express a particular subject, the one may repeatedly correct the statements written by the other and vice versa, until perhaps a compromise or synthesis emerges—which may have been proposed by one or more third parties. This is variation and selection at work: different people contribute different text fragments, some of which are clear, accurate and relevant, some of which are less so. The continuing process of revisioning by a variety of users will normally leave the good contributions in place, and get rid of the poor ones, until the text as a whole provides a clear, coherent and in-depth coverage of its subject, without glaring mistakes. When the subject is controversial, an evolved text will typically provide a balanced overview of the different perspectives, noting the arguments pro and contra each position.

This example shows the true power of stigmergy: thanks to the availability of the medium (in this case the Wikipedia website) independent agents together perform a complex activity that is beneficial to all, minimizing social frictions and stimulating synergy—and this without need for a hierarchical control or coordination, a clear plan, or even any direct communication between the agents (Heylighen 2007b). In the present web, similar mechanisms are being used to collaboratively develop not just an encyclopedia of existing knowledge, but a variety of novel knowledge and applications, including various types of open source software, scientific papers, and even forecasts of the world to come (using web versions of the well-known Delphi procedure). Thus, Internet-supported stigmergy strongly promotes the collective development of new knowledge and tools.

Again, there is a direct analogy with the functioning of the brain. Whereas quantitative stigmergy can be likened to the neural processes that characterize subconscious cognition, qualitative stigmergy is most akin to the higher-order, symbolic processes that we associate with conscious thought. According to evolutionary psychology, the brain consist of an array of many, largely independent modules, each specialized in a particular task—such as recognition of specific shapes, emotions, or control of specific movements—that work in parallel. These brain modules have few direct connections that allow them to communicate so as to form a global picture of the situation. One way for them to pool their expertise is by exteriorizing the inferences made by some of the modules, so that the results can be perceived, i.e. re-entered into the brain and thus processed by the other modules. Exteriorizing cognition takes place through the creation of physical symbols, such as drawings, utterances or writings, that represent the mental contents. Typical examples of this process are talking to oneself, or taking notes and drawing schemas while thinking about a complex problem. This is an example of stigmergic interaction between the modules within one's brain: a module's outcome through action is converted into a change of the environment; this change is then perceived again, triggering new inferences by the same or other modules, that produce a new action, and a subsequent modification of the external symbols; in this way, an idea is step by step elaborated and refined.

As the individual becomes experienced with this process, however, shortcuts are developed and symbols are interiorized again. Thus, children talking to themselves while thinking will soon learn to use inner speech, i.e. forming sentences in their head without actually vocalizing them. According to the global workspace theory (Baars 1997), higher-level consciousness is nothing more than the "working memory" or "theater" within the brain where these interiorized symbols are produced and combined, so that they can be submitted to the scrutiny of the various more specialized modules. This global workspace is a shared internal environment or mediator that the brain has evolved in order to facilitate the coordination and control of its otherwise largely autonomous and instinctually reacting modules. From this perspective, it seems obvious that the world-wide web too is a global workspace that serves the coordination of autonomous individuals, together forming a brain-like system at the planetary scale. The novel ideas developed collectively in that workspace form the equivalent of the thought processes of the global brain. As the conventions, protocols, software and hardware tools supporting this workspace evolve, they become more efficient, and novel ideas and solutions to existing problems will be produced more quickly and more easily, thus greatly increasing collective intelligence and creativity.

## The dynamics of global evolution

The model proposed in this paper sees evolution characterized by unambiguous advance or progress towards more synergetic systems, characterized by reduced friction, and therefore more productive use of resources. The same kind of progress can be found on the levels of matter, energy, information, cognition and cooperation, thus spanning the whole hierarchy from physics to society (Heylighen 1999a, 2007a; Stewart 2000). The technological and institutional innovations come together in what we have called the "medium", i.e. the part of the environment shared by different agents that is used as their workspace, or means for communication and collaboration. The thrust of the argument is that there is no need for intentional use or design of such a workspace: any medium that can accumulate changes produced by the agents tends to evolve into a mediator that coordinates their actions, and thus promotes synergy. This is because the variation-and-selection dynamics that underlies individual evolution is extended to collective evolution via the mechanism of stigmergy, where the changes to the medium made by one agent indirectly affect the actions of the other agents.

### The socio-technological singularity

A major effect of stigmergy is the *acceleration* of evolution: a solution to an evolutionary problem found by one agent can now, by impressing it upon the medium, be used and improved by other agents. Since the medium benefits all agents' fitness, there will be a selective pressure on the agents to find solutions that make the medium itself more powerful. The further the medium extends, and the easier it becomes for agents to interact with it, the quicker innovations will spread and undergo further improvements. This leads to a self-reinforcing process: improvement of the medium facilitates further innovation, which in turn helps improve the medium. This explains the explosive advance in science and technology over the past centuries, as exemplified by the (at least) exponential increase in the number of scientific publications (Kurzweill 2006).

Some aspects of such accelerating growth can be captured in mathematical models. An elegant example is the explanation by (Korotayev et al. 2006; Korotayev, 2007) of the hyperbolic growth of the world population until 1960 (von Foerster et al. 1960). In the model, population growth is initially modeled by a traditional logistic growth equation (1), where population $N$ starts by growing exponentially but then slows down until it reaches the maximum value expressed by the carrying capacity $bT$ of the environment. This carrying capacity is proportional to the overall productivity $T$ of technology, i.e. its ability to extract from the natural environment the resources necessary for survival. In a second equation (2), the growth $dT$ of technological productivity is considered to be proportional to the technology *T* that is already there (simple exponential growth), but also to the population number *N*, under the simple assumption that more individuals will discover more innovations.

$$dN/dt = a\,(bT - N)\,N \qquad (1)$$

$$dT/dt = cNT \qquad (2)$$

The authors show that the two equations together produce a hyperbolic growth curve that mimics the observed historical growth of world population with a surprising accuracy (explaining over 99% of the variation for the period 500 BC–1962). The growth is much faster than could be expected from a traditional logistic or even exponential curve, because of the positive feedback between population ("agents") and technology ("medium").

It is obvious that hyperbolic growth (which would lead to an infinite value within a finite time) is not sustainable for population. This explains why the model breaks down after 1962 when the demographic transition to smaller family size starts to occur. But that can be perfectly understood from an evolutionary point of view, as a shift from r-selection (fast reproduction, short life) to K-selection (slow reproduction, long life) (Heylighen & Bernheim 2004). When life becomes less risky, fitness is better served by long-term investments in longevity and quality of life than by a short-term strategy for producing as much offspring as possible. The most recent population models of the UNDP therefore forecast a stabilization of world population by about the year 2100.

However, this does not entail an end to the population-technology feedback: K-selection implies an on-going growth in the general health, wealth, development, education and even IQ levels (Heylighen & Bernheim 2000a, 2004) of the population. This in turn will increase the potential of each individual to innovate, and thus the speed of technological progress. Vice versa, technological progress enhances the general development level of individuals, via improvements in health, education, wealth, autonomy, etc. Thus, it is conceivable that the technology-supported hyperbolic growth in human population has simply shifted to a growth in "human potential" or "human development". While human potential may be difficult to quantify, the human-technology feedback implies that technology too should obey a hyperbolic growth, and this could be measured via various indices of productivity.

The essence of hyperbolic growth is that it will produce an infinite value within a finite time. In mathematics, the point where the value of an otherwise finite and continuous function becomes infinite is called a *singularity*. It can be seen as the point where quantitative extrapolation must break down. Vinge (1993), Kurzweil (2006) and others have argued that technological innovation is racing towards such a singularity. In the short term, scientific and technological innovation appears to obey an exponential growth, as illustrated by a variety of statistical trends (e.g. Moore's law, or the increase in scientific publications). In the somewhat longer term, however, the rate of growth itself appears to growing (Kurzweil 2006). For example, our review of information transmission speeds over the past two centuries indicates a much faster than exponential growth. This makes the process super-exponential, and possibly hyperbolic. Extrapolation of these trends leads to different estimates for the year of the singularity:

- 2005 to 2030 according to Vinge's (1993) interpretation of the increase in machine intelligence;
- 2026 according to the original extrapolation of hyperbolic population growth by von Foerster et al. (1960);
- 2045 according to Kurzweil's (2006, p. 136) compilation of technological trends;
- 2052 ± 10 according to Johansen and Sornette's (2001) extrapolation of various population, economic and financial growth curves.

The implication is that at some point within the next half century the speed of innovation would—at least for all practical purposes—become infinite. This means that an infinite amount of knowledge (and the concomitant wealth) would be generated in a finite time. At such a point, every further extrapolation that we could make based on our present understanding of evolution, society or technology would become meaningless. The world will have entered a new phase, where wholly different laws apply. Whatever remains of the global system as we know it will have changed beyond recognition.

While my interpretation of accelerating change is somewhat more cautious than the one of "singulitarians" like Kurzweil, I believe that this acceleration does point to the evolutionary emergence of a higher level of organization—what Turchin (1977) has called a *metasystem transition*. An example of such an evolutionary transition is the emergence of multicellular organisms from individual cells. The *global superorganism* (Stock 1993; Heylighen 2007c) directed by its global brain (Heylighen 2004) is a metaphor for the "metasystem" that would be formed in this way—a system that would integrate the whole of humanity together with all its supporting technologies and most of its surrounding ecosystems, and that would function at a level of intelligence, awareness and complexity that we at present simply cannot imagine.

Is global progress cyclic?

As part of a volume where many contributions have their roots in the "long wave" tradition (e.g. Modelski 2006; Devezas & Modelski 2003), this paper should also address the issue of periodicity in the evolution of the world system. It is clear that in a model where the focus is on ever accelerating growth racing towards an apparent singularity, there is little room for slow oscillations, i.e. long-term cycles of renewal and growth followed by decline and fall, which repeat at regular intervals. Yet, as illustrated by Kurzweil (2006, p. 43) and Johansen & Sornette (2001) it is possible to superpose a certain amount of cyclicity on an exponential or hyperbolic growth curve, by assuming that the speed (or acceleration) of growth oscillates somewhat around its "normal" value. A rationale for doing this is the assumption that innovation is not a continuous process, but a sequence of discrete discoveries, inventions or paradigm shifts, each bringing forth a new technology, institution, or way of thinking. Each major innovation needs time to diffuse and be adopted by the whole of society. Such diffusion process is traditionally modeled as a logistic or S-curve, characterized by an initially fast growth which then slows down until saturation, when it has reached most of the population.

To achieve periodicity, we moreover need to assume that major innovations do not occur independently, but that the later one "waits" before it starts spreading until the former one has reached saturation. This is a not unreasonable assumption, if we consider major socio-technological paradigms that dominate people psychologically and economically to such an extent that they are not interested in exploring alternatives until the present paradigm has run out of steam. However, when we consider the variety of smaller and larger innovations that are constantly being produced by evolutionary trial-and-error, some of which depend on or compete to various degrees with others, then the fixed "waiting period" between innovations becomes less plausible.

Moreover, even if innovations were polite enough to wait until their predecessor has run out of steam, the time to saturation is unlikely to be constant. Some inventions are simply more difficult to adopt than others, because of psychological, economical or infrastructure constraints. For example, the world-wide web and cellular phone technologies appeared at about the same time, but the latter spread much more quickly than the former (at least outside the USA), because portable phones are less costly and easier to learn using than Internet-connected computers.

Finally, even if we could determine a "typical" delay for major inventions to diffuse, we would find that delay to be decreasing because of accelerating progress (Kurzweil 2006, p. 43). Thus, we could hope to find a superposition of increasingly short logistic curves on top of our overall super-exponential curve pointing towards infinity. (A related, but simpler mathematical model is Coren's (1998) "logistic escalation": a sequence of ever shorter and steeper logistic growth processes culminating in a singularity.) However, if we take into account the variability in size and diffusion speed of innovations, the net result is likely to be an almost random seeming pattern of fractal fluctuations with increasingly high frequencies around the large-scale super-exponential trend.

While this analysis explains my skepticism towards "long wave" models of the present process of globalization, it does not imply that I wish to dismiss these models altogether. The intensity of present-day acceleration implies that in comparison during most of history progress occurred at an almost glacial pace. Centuries ago, technological advance was slow enough that it may not have been noticeable within one generation. Moreover, the number of simultaneous inventions was much smaller.

In such circumstances, the speed of diffusion is likely to have been practically constant, and there would have been much less competition between parallel innovations. This would make the above assumptions leading up to periodicity much more plausible.

**Conclusion**

In this paper I have reviewed the evolutionary mechanisms that drive the present process of globalization. Evolution in general is a trial-and-error learning process, leading to the progressive accumulation and improvement of knowledge. Its subjects are agents, which I defined as cybernetic systems that act upon their environment in order to achieve their goals. Natural selection of agents and the knowledge they use to plan their actions pressures them to become progressively better in achieving their goals.

In the realm of technology, this progress is most visible as ephemeralization, the on-going increase in "total factor productivity". It can be conceptualized most simply as a reduction in the friction that normally produces the dissipation of energy, information and other resources. As a consequence, ever more results can be achieved with ever fewer resources. On the level of society, this entails a spectacular expansion in the flows of matter, energy and information that circulate across the globe. Thus, the obstacles of time, distance and material scarcity have largely vanished, making the different parts of the world increasingly interconnected.

Connectivity implies an increase in the number of agents one is interacting with, and therefore an increase in social complexity, with the concomitant threats of competition, conflict, and confusion. These problems too can be conceptualized most generally as a form of friction, i.e. the (generally unintended) obstruction of one agent's actions by one or more other agents' actions. As in the case of technological progress, the trial-and-error of evolution will tend to reduce this social friction by creating adapted institutions. Institutions, or more generally *mediators*, are systems that coordinate the activities of different agents so as to minimize friction and maximize synergy.

A largely overlooked, but very powerful, mechanism for the spontaneous evolution of mediators is *stigmergy*, which relies on the medium or environment shared by the agents. Stigmergic interaction means that the change produced by one agent to the medium stimulates another agent to perform a complementary action, promoting their collective benefit even without any conscious intention to cooperate. Stigmergic activity will gradually reshape the passive medium into an active mediator, which elicits and directs the agents' actions.

The most "ephemeralized" example of the technological infrastructure underlying global connectivity is the Internet. A quick inspection shows that it provides a near ideal medium for stigmergic interaction (cf. Heylighen 2007b): it instantaneously connects people all across the planet, is nearly always and everywhere available, can be used virtually without cost, is plastic enough to accommodate practically any "shape" or information that is imprinted upon it, while it will accurately register and store this information for as long as necessary. What it still lacks are the more evolved mediator functions. Yet, there already exist several examples of Internet services—such as the Google search engine or the Wikipedia website—that very successfully apply stigmergic principles to coordinate individual activities, thus offering their users a form of distributed intelligence well beyond the capabilities of a single individual.

A straightforward extrapolation of this evolution that injects ever more intelligence into the Internet leads me to expect a near-term shift from World-Wide Web to Global Brain (Heylighen & Bollen 1996). The "Global Brain" is more than a fancy term for a large-scale intelligent system, though: the analogy runs much deeper. An analysis of the stigmergic mechanisms that seem most effective in supporting such distributed intelligence shows that they are virtually identical to the mechanisms used by the human brain. The quantitative stigmergy exemplified by "ant algorithms" is nearly identical to the process of Hebbian or reinforcement learning that differentially strengthens connections between neurons in the brain. The "ants" that trace and explore the quantitatively weighted network formed in this way correspond to human or software agents searching the web, or to bursts of activation spreading across the brain. Qualitative stigmergy, which is the true motor of innovation, can be seen as the basis of symbolic consciousness in the brain. It is exemplified on the web by a variety of collaborative, "open access" sites where people freely improve on each other's contributions (Heylighen 2007b).

The paper concluded with an attempt to provide a quantitative underpinning to this purely conceptual forecast of the emerging global network. Over the past centuries, both technological and social evolutions appear to be accelerating spectacularly. The stigmergic interaction between medium (technology) and agents (society) moreover points to a positive feedback relation, where the one catalyzes the development of the other. Such cross-catalytic interaction is elegantly captured by the mathematical model proposed by Korotayev et al. (2006) to explain the growth of world population. However, the resulting hyperbolic growth model entails a singularity, i.e. a point in the near future where the speed of innovation becomes virtually infinite. Different authors have estimated such a singularity to take place around the year 2040, give or take a decade or two. While I do not want to put too much emphasis on such a number, which I consider to be much less reliable or important than the qualitative transition that it represents, these number agree with my intuition that a momentous change is likely to happen within a surprisingly short term—probably still within my own lifetime.

The speed and radicalness of the transition, and the inscrutability of what will come after, implies that I have little confidence in traditional methods of quantitative extrapolation, and in particular in those based on "long wave" periodicity. While I see a possible utility in distinguishing some degree of cyclicity in the long-term upward trends that I have described (cf. Johansen & Sornette 2001), the extreme acceleration of change implies that those cyclical fluctuations can only become shorter and shallower as the metasystem transition to the Global Brain regime approaches.

As to a more qualitative extrapolation of social and technological trends, I refer to a slightly older companion paper which reviews the cybernetic organization and evolution of the emerging global "superorganism" (Heylighen, 2007c). In spite of the intrinsic difficulty of forecasting an evolution that is so rapid, complex and radical, I hope that both papers together may offer a reasonably realistic outline of the momentous transformations that our globally networked society is undergoing.